# ALL NEAREST NEIGHBOUR CALCULATION BASED ON DELAUNAY GRAPHS

Nasrin Mazaheri Soudani [1], Ali Karami[2]

[1]Department of Computer Engineering, University Of Sheikhbahaiee, Isfahan, Iran
n.mazaheri@shbu.ac.ir
[2]Department of Computer Engineering, University Of Sheikhbahaiee, Isfahan, Iran
ali.karami@shbu.ac.ir

## ABSTRACT

*When we have two data sets and want to find the nearest neighbour of each point in the first dataset among points in the second one, we need the all nearest neighbour operator. This is an operator in spatial databases that has many application in different fields such as GIS and VLSI circuit design. Existing algorithms for calculating this operator assume that there is no pre computation on these data sets. These algorithms has o(n\*m\*d) time complexity where n and m are the number of points in two data sets and d is the dimension of data points. With assumption of some pre computation on data sets algorithms with lower time complexity can be obtained. One of the most common pre computation on spatial data is Delaunay graphs. In the Delaunay graph of a data set each point is linked to its nearest neighbours. In this paper, we introduce an algorithm for computing the all nearest neighbour operator on spatial data sets based on their Delaunay graphs. The performance of this algorithm is compared with one of the best existing algorithms for computing ANN operator in terms of CPU time and the number of IOs. The experimental results show that this algorithm has better performance than the other.*

## 1. INTRODUCTION

The all nearest neighbour is a kind of operator that takes two datasets as input and for each point in first dataset, finds nearest neighbour from second dataset. This operation has many applications in different fields. Its applications are clustering, spatial data mining, image processing, geographical information system, urban planning, resource allocation problem, mobile data based, graph based computational learning, simulation and VLSI design. [4-14]

The existing methods for calculating ANN operator assumes that there is no pre computation on these data sets. These algorithms have O(n*m*d). With assumption of existence some pre computation, an algorithm can be given for calculating it that its cost is much lower than when for each point, nearest neighbour algorithm is used independently because there are many repetitive actions in finding nearest neighbours and these actions can be done overlapped or the information that obtained from recent nearest neighbour can be used for finding next nearest neighbour[8].

The new kinds of index structures were presented in some papers that had good performance for ANN but because of these indexes are provided only for calculating this type of operator and their performance hasn't been investigated for other operators, these index structures aren't popular and common[15-17]. There are many other algorithms based on tree indexes such as R-trees[16]. Because of R-tree and many other kind of common tree index structures are based on rectangular nodes and don't have accurate information about nearest neighbouring relationship of nodes (even with various pruning metrics that introduced to improve performance) algorithms based on these structures don't have good performance for ANN operator.

One of the popular and common pre computations in multidimensional datasets is voronoi diagram. This diagram divides space into disjoined regions based on a dataset such that a region related to a point is the locus of all points in space that this point is nearest neighbour of them. The dual of this diagram is delaunay graph that its vertices are points in the dataset and there is a link between each two points which their voronoi regions are adjacent. A voronoi diagram for a dataset with n points can be constructed in $\Omega(nlogn)$. For useful and accurate information which voronoi diagrams are provided, they are very popular and common and as had been stated in [18] they have good performance for computing common operators in multidimensional data[18].

As our knowledge, up to now, there is no algorithm for calculating ANN operator in the general case with assumption of being voronoi diagrams on datasets. The only existing algorithm supposed both datasets are the same and for all point in the dataset finds nearest neighbour from the same dataset[18].

In this paper an algorithm is provided for calculating ANN operator. It is supposed that there are different voronoi diagrams on both datasets. According to popularity of voronoi diagrams this assumption seems reasonable. In this algorithm at first one point of the first dataset is selected and for this point the delaunay graph of the second dataset traversed depth first until a point is found that the point of the first dataset is located on its voronoi region. In the depth first search of the second graph, every time the neighbour is selected as the next point that has less distance to the point of the first dataset related to all other neighbours. So a kind of objectively is seen in the depth first search of the second dataset.

By finding the first nearest neighbour, a pair of points from the first and the second datasets is found that are located near each other. From this time, for all other points in the first dataset the same process is repeated except that instead of beginning depth first search of the second dataset from random point in this set, this search is started from a point that probably is located near the point of the first set so for each point in the first set, nearest neighbour can be found with less steps in depth first search of the second dataset.

Considering the useful information about nearest neighbouring which voronoi diagrams and delanuay graphs provide, in this algorithm, the information obtained from finding recent nearest neighbour is used for searching next nearest neighbour and a kind of intelligence and objectivity is used for finding nearest neighbours. This algorithm needs no additional preprocess on voronoi diagrams such as sorting and work with common type of voronoi diagrams. The performance of this algorithm in terms of CPU time and the number of IOs has been compared with one of the best recent presented algorithms and in most cases the introduced algorithm has shown better performance.

In the rest of this paper, in section one voronoi diagrams and delaunay graphs are introduced with more details, in section two previous works about this operator and join operators in multidimensional data have are probed and general aspects of them are compared, in section three the main algorithm is presented, in section four validation of this algorithm for calculating ANN operator is proven, in section five the performance of this algorithm is investigated and finally in section six general conclusion of this paper is discussed.

## 2. PRELIMINARIES

The formal definition of the ANN operator according to [1] is seen in below.

$$\forall (r, s) \in R \underset{k\text{-}nn}{\bowtie} S, \forall (r, s') \in R \times S \setminus R \underset{k\text{-}nn}{\bowtie} S : \| r - s \| < \| r - s' \| \ .$$

Two input datasets in this definition named R and S and ANN calculates for each point in R its nearest neighbour in S.

Voronoi diagram is a way for dividing space into separate regions according to a dataset which in it each region is allocated to a point in the dataset. The Region for a point is called site or voronoi cell and is the locus of all points of space that the distances of them to this point are less than all other points in the dataset. The voronoi cell of a point is a covex polygon and each border of it is a part of perpendicular bisector of this point and another point in the dataset around this point. All points around a point in the dataset that have a common border to this point are called voronoi neighbours. delaunay graph is dual of voronoi diagram and in it the points of the dataset are considered as the vertexes and there are edges between two points if and only if two vertices are neighbours to each other

## 3. RELATED WORK

The simplest way for calculating ANN operator is using a nearest neighbour operator for each point in the first dataset. This approach is called nested loop and in it different ways for calculating nearest neighbour can be used. In this approach if there is an index on each two dataset, regardless of the index structure, it can be used.

In many approaches based on nested loop it is supposed that the second dataset and its index are in main memory. So in cases that the second dataset is large, they can't be used. In some approach such as [8] tries that query points that are near to each other are retrieved sequentially for finding nearest neighbour because a large percentage of memory pages that are used for finding a nearest neighbour are the same pages that retrieved for finding previous nearest neighbour and so they will be in LRU memory and retrieving of them has lower cost. The main difference this approach and the approach presented in this paper is that, the algorithm of this paper uses the information provided from previous nearest neighbour heuristically and doesn't rely on memory techniques.

The algorithms presented so far for ANN can be divided into three categories[8]. In the first category it is supposed that there is an index on each data set. The algorithms in this category often use the tree indexes such as R trees or introduce new kinds of index structures[16]. In this algorithms both indexes are traversed simultaneously and all nodes that are not promised are pruned and finally the nearest neighbour of each leaf node found by searching in promising nodes. most algorithms that are based on R tree because of the query point's R tree is traversed only one time use queues in nodes of data tree for holding nodes of the query tree that are promised for finding nearest neighbour of them. The main problem of these algorithms is that they need large memory for holding and updating these queues. The algorithms that are based on new index structures are less popular because of these structures are not conventional. These structures often act well for this operator but their actions for other usual operators have not probed in these papers.

According to an analysis provided in [15], the search for finding nearest neighbour of a point can be divided into two phases. In the first phase, we arrive to the neighbourhood of query point and in second phase the nearest neighbour of the query point is found by searching the regions that are in the neighbourhood of the query point. Tree based indexes such as R trees are successful in first phase but don't act well for the second phase. In this phase for finding actual nearest neighbour all nodes that are knows promising have to be probed. Although many measures have been introduced for pruning nodes but because of rectangular structure of R tree nodes, these nodes have not accurate information about nearest neighbouring and so the number of promising nodes for a query point will be high.

The voronoi diagrams don't act well in first phase but because they have accurate coordinates of the nearest neighbouring region related to each point they acts well in second phase. So if the search for finding nearest neighbour of a point in ANN is started from neighbourhood region of that point, nearest neighbour search with the help of voronoi diagrams can be done faster than tree indexes and it doesn't need to frequently repeat the first phase which voronoi diagrams are weak on it.

In [18], a solution for computing nearest neighbour operator with assumption of being voronoi diagram on dataset is presented. In this solution, the horizontal lines that pass through the vertex of the diagrams divide space into strips and then with the help of binary search in o(log n), the strips that contains query point is determined. In this step by another binary search in this strip nearest neighbour of query point can be found. This algorithm has good time complexity but the disadvantage of this approach is that high pre computation should be done on voronoi diagram and its runtime memory usage is high.

In [18] an approach for calculating ANN operator with the help of voronoi diagrams is presented. In this book, the specific type of ANN operator is considered which in it supposed that the two datasets are equal and there is a voronoi diagram on it. In presented algorithm for each point of voronoi diagram, its distances from all voronoi neighbours are calculated and the voronoi neighbour that has less distance is considered nearest neighbour of that point. This algorithm is not suitable for cases that two datasets are not equals.

In the second category of algorithms it is supposed that only one dataset has index. In these algorithms a common method is to build an index structure for another dataset and apply one approach presented for the first category[19]. In [20] a method for this category of problem based on hash presented which in it the space is partitioned based on the nodes of the existing index and the points are mapped into partitions.

In the third categories of algorithms it is supposed that none of dataset has index. The algorithms presented with this assumption are usually based on hash method[21, 22]. In these algorithms, the space is partitioned into disjoint regions and the points of both data sets are mapped into the regions. then all points from both datasets that are in a region, are loaded into main memory and with the help of plane sweep method, are probed for finding nearest neighbours. If the data related to a region don't fit in memory, this region is divided to smaller regions and each region probed separately[8].

## 4. THE MAIN ALGORITHM

### 4.1. Assumption

In this algorithm it is supposed that there is a voronoi diagram for each dataset. voronoi diagrams are stored in flat files. This means that for any point, its voronoi neighbours (the adjacency list of delaunay graph) and voronoi region is stored. Reading information about a point and determining its voronoi neighbours is considered as one access to a file and one IO and the goal of this algorithm is to reduce the number of access to the files.

### 4.2. Nearest neighbour algorithm

For finding the nearest neighbour of a query point among the points of a dataset that there is a delaunay graph on them, this graph is traversed with the help of an extension of depth first search algorithm. According to depth first, the search for finding nearest neighbour is started from an arbitrary vertex of the graph and a stack is used for holding the vertexes that may expand in future for exploring their children. If this vertex is the nearest neighbour of the query point (this means that the query point is located in voronoi region of this vertex), search is finished else this vertex is marked as visited point and its voronoi neighbours that has not been marked yet, are pushed into the stack in order of decreasing their distances to the query point. In the next step, the top vertex is removed from the stack and if it has not been marked so far, it is expanded and probed like the previous vertex.

Due to the special order which was observed by adding nodes to the stack, the top nodes is closer than all other nodes in the stack to the query point. So there is more hope that from this point, the nearest neighbour point is reached sooner. Therefore by this algorithm, we get closer to query point step by step from the initial point. Indeed in each step we go from recent point to a point that gets us closer to the query point.

In each step of this algorithm, the A* heuristic search algorithm[23] is used for selection the next point among voronoi neighbours of recent point. In A* algorithm, the cost of going from a point X to the goal is estimated by $F(X)= h(X)+g(X)$ which in it $g(X)$ is the cost of going from starting node to the current node and $h(X)$ is a heuristic estimate of the distance from X node to the goal.

In the nearest neighbour problem, the goal is the nearest neighbour of the query point and the nearest neighbour hasn't been found yet but is probably near the query point so the query point is considered as goal. Each step needs to retrieve one block of voronoi diagram related to a point from file and has an IO. So $g(x)$ is the cost of retrieving one block and for all voronoi neighbours is the same. $h(X)$ should estimate the number of steps from candidate point to the goal. The Euclidean distance to the query point is considered as $h(X)$ in our algorithm because whatever the distance to the query point is lower, the steps for reaching query point is probably lower too.

If the density of data point is uniform everywhere in the dataset, this estimation for $h(x)$ is good but if the area between the starting point and the nearest neighbour has high density, $h(x)$ is bad estimation for the number of steps to the nearest neighbour. Because the ANN algorithm presented in this paper calls nearest neighbour algorithm (except the first time) so that the starting point and the query point are approximately near each other, this estimation acts well for all cases.

If the heuristic h satisfies the additional condition $h(x)<= d(x,y)+h(y)$ for every edge x, y of the graph (where d denotes the length of that edge) in A* algorithm, then h is called monotone and A* can be implemented more efficient ( no node needs to be processed more than once ). In our algorithm because the Euclidean distance from each point and the query point is considered as $h(X)$ and in Euclidean space the length of straight line is the shortest distance between two points, $h(X)$ satisfies this condition and A* search acts like Dijkstra algorithm.

### 4.3. All nearest neighbour algorithm

If the first and the second datasets are called as Q and P, the ANN operator should find the nearest neighbour of each point in P among the points of Q. for this purpose the graph of Q traversed by an expansion of depth first search and for each point the NN algorithm is called. The main advantage of this algorithm is that for each step NN algorithm is not started from a random point in P but it is started from a point that is probably in neighbourhood of the point that we want to find its nearest neighbour and therefore by fewer steps its nearest neighbour can be found.

at first one point of Q is selected randomly and for this point the NN algorithm is called with starting from another random point of P. so the nearest neighbour of Q point is calculated. The Q point is marked as visited and voronoi neighbours of it that has not been marked as visited yet are added to a stack in order of decreasing their distance from the recent found nearest neighbour.

In next step the top element of stack is popped and if it has not been visited yet, its nearest neighbour is calculated. For finding its nearest neighbour the NN algorithm is called and started from the point that was nearest neighbour in previous step. Due to special order that we add nodes to the stack, the top element of stack that is explored now is closer than all other voronoi neighbours to previous nearest neighbour and these two points are probably in neighbourhood of them. So with starting search from previous nearest neighbour we can get to new point's nearest neighbour with fewer steps.

Indeed by finding the first nearest neighbour in this algorithm we reach to the points from two datasets that are in neighbourhood of them. This is used for finding next nearest neighbour to decrease the number of steps. If the first point in Q and its nearest neighbour are considered as two pointers in the space that are in neighbourhood of them, in next step the pointer of Q moves and goes to the point that has lower distance to the pointer of P. next the pointer of P moves few steps by calling NN algorithm and locates near the Q pointer. So two pointers move near each other, traversed all space and find all nearest neighbours.

## 5. VALIDATION OF THE ALGORITHM

The delaunay graphs are connected graphs. So with depth first search all nodes of them visited once. The ANN algorithm use depth first search for traversing the point of Q dataset. So all nodes of Q are visited and the NN algorithm is called for them. in NN algorithm of a point the delaunay graph of P dataset is traversed until a point is found that its voronoi region contains query point. Q and P datasets are in the same space. The voronoi diagram of p divides this space into disjoint regions. These regions cover all space and the Q point that NN called for it is in this space so it is locate definitely in one region. Because the depth first search traverses all points in P and their regions the region that contains the Q point are explored and the nearest neighbour is found absolutely.

## 6. PERFORMANCE EVALUATION

For evaluating this algorithm, it is compared with the most popular algorithms based on Rtree[16] that to our knowledge has the best performance compared with all other algorithms for ANN operator in the general case. The performance of these two algorithms is compared in terms of CPU times and the number of IOs.

Two main parameters of ANN operator are the number of points in the data set and the number of points in the query set and the performance of algorithms are compared with each other based on these parameters. The primary evaluations are done in two primary phase. In each phase the evaluation changes are probed with maintaining one parameter and changing another. The data needed for evaluating are both real data from [24] and randomly generated data with uniform distribution in space.

Two ANN algorithms, Voronoi diagram and R Tree have been implemented with java programming language in NetBeans environment. The block length for both RTree and Voronoi were 1Kbytes. The evaluations were performed in Intel 2 corer 3.06 GH with 2 Gbytes of RAM and 150 Gbytes of HDD. it is supposed that Voronoi and Rtree indexes are in hard and their blocks loaded into main memory on demand.

The results of performance evaluation are seen in below figures. In figures 1-2 the changes of CPU time and the number of IO based on changes in the number of query points is seen. The number of data points was 10000 and the number of query points is changed from 5000 to 24000.

In figures 3-4 the CPU time and the number of IO VS changes in the number of data point is seen. The number of query points is 20000 and the number of data points is changed from 5000 to 24000.

As you can see in all figures the voronoi based algorithm presented in this paper has better results compared with voronoi based algorithm in most cases.

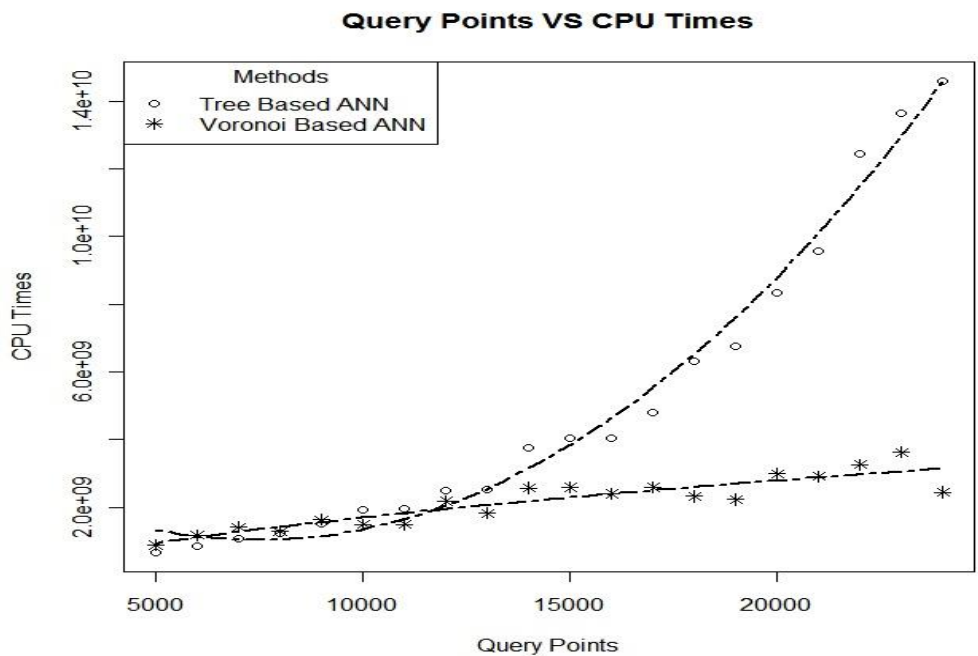

**Figure 1**: **The CPU times related to the number of query points**

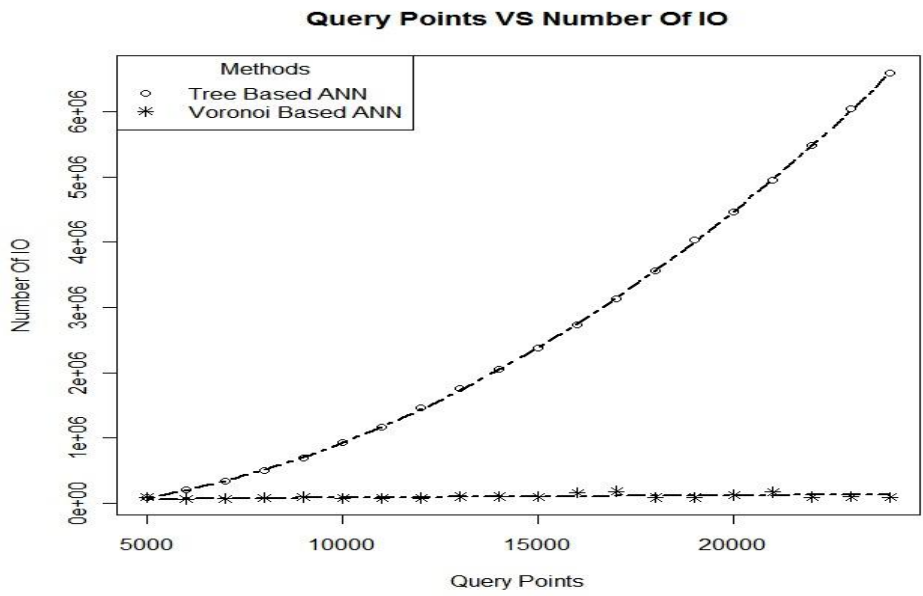

**Figure 2**: The number of IOs related to the number of query points

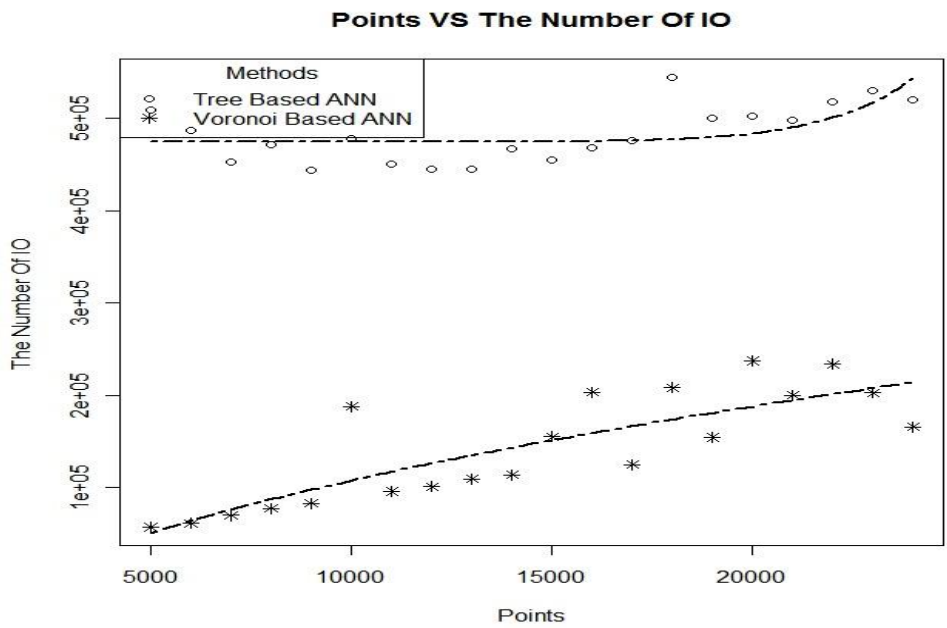

**Figure 3**: The number of IOs related to the number of data points

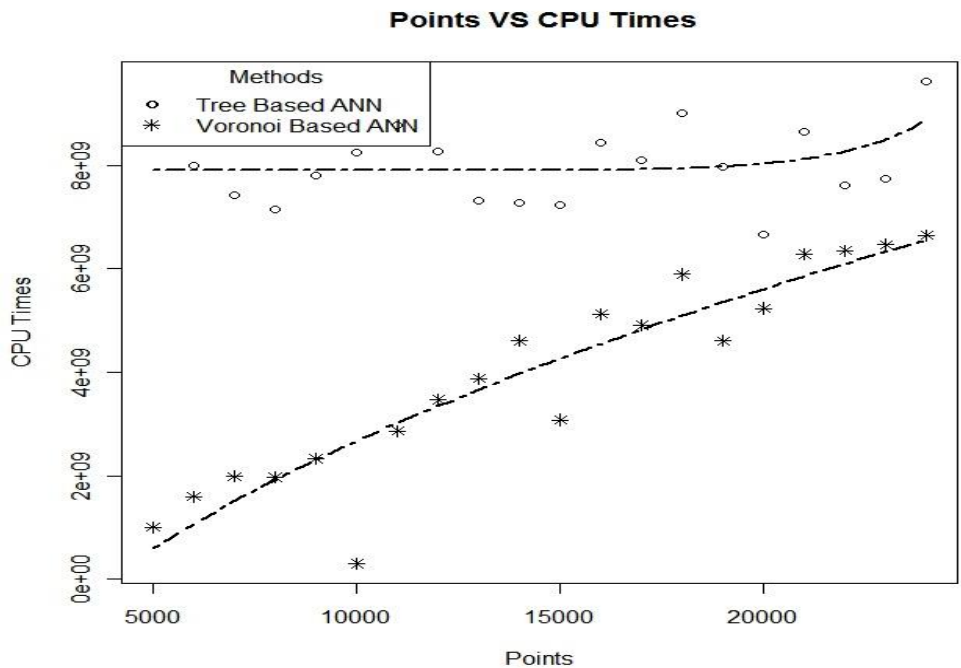

**Figure 4**: The CPU times related to the number of data points

## 6. CONCLUSION

A new algorithm for computing ANN operator on spatial data sets is presented in this paper. This algorithm assume that there are precomputed voronoi diagram and delaunay graphs on the data sets. This assumptions are reasonable because of the popularity and usefulness of the voronoi diagrams. In this algorithm, we first compute the nearest neighbour of a random point in the first data set among points of the second data set with a depth first search. Then, we search and examine neighbours of these two points for finding other pairs of nearest neighbours. New found pairs are also used for finding another pairs. These routine is continued until all nearest neighbours are calculated. This algorithm uses voronoi diagrams as indexes for finding points around found pairs of nearest neighbours.

The performance of this algorithm is compared with one of the best existing algorithm based on R indexes. These experimental comparison shows that the presented algorithm has lower CPU time and the number of IOs than the other.